# Influence of Material Parameter Variability on the Predicted Coronary Artery Biomechanical Environment via Uncertainty Quantification

Caleb C. Berggren[1†], David Jiang[1†], Y.F. Jack Wang[1†], Jake A. Bergquist[1,2,3], Lindsay C. Rupp[1,2,3], Zexin Liu[2,4], Rob S. MacLeod[1,2,3], Akil Narayan[2,4], Lucas H. Timmins[1,2,5,6]

[1] Department of Biomedical Engineering
University of Utah
Salt Lake City, UT, USA

[2] Scientific Computing and Imaging Institute
University of Utah
Salt Lake City, UT, USA

[3] Nora Eccles Cardiovascular Research and Training Institute
University of Utah
Salt Lake City, UT, USA

[4] Department of Mathematics
University of Utah
Salt Lake City, UT, USA

[5] School of Engineering Medicine
Texas A&M University
Houston, TX, USA

[6] Department of Biomedical Engineering
Texas A&M University
College Station, TX, USA

† The first three authors contributed equally to this work.

Correspondence:
Lucas H. Timmins, Ph.D.
School of Engineering Medicine
Texas A&M University
1020 Holcombe Blvd.
Houston, TX 77030
lucas.timmins@tamu.edu

Phone: (713) 677-7491

ORCID ID:
C.C. Berggren – 0000-0002-8770-9690
Y.F.J. Wang – 0009-0007-2677-2091
J.A. Bergquist – 0000-0002-4586-6911
R.S. MacLeod – 0000-0002-0000-0356
A. Narayan – 0000-0002-5914-4207
L.H. Timmins – 0000-0002-8707-8120

**Acknowledgments**

This work was supported, in part, by the National Institutes of Health grants R01 HL-150608 (L.H.T.), U24 EB-029012 (R.S.M., A.N.), P41 GM-103545 (R.S.M.), R24 GM-136986 (R.S.M.), American Heart Association grant 23PRE1019455 (C.C.B.), National Science Foundation GRFP (L.C.R.), and the Nora Eccles Treadwell Foundation for Cardiovascular Research (R.S.M.). The Authors thank Prof. Jeff Weiss and Dr. Steve Maas for their insightful discussions and assistance with the `FEBio` Software Suite, and Drs. Habib Samady and David Molony for access to the clinical data.

**Abstract**

Central to the clinical adoption of patient-specific modeling strategies is demonstrating that simulation results are reliable and safe. Indeed, simulation frameworks must be robust to uncertainty in model input(s), and levels of confidence should accompany results. In this study, we applied a coupled uncertainty quantification-finite element (FE) framework to understand the impact of uncertainty in vascular material properties on variability in predicted stresses. Univariate probability distributions were fit to material parameters derived from layer-specific mechanical behavior testing of human coronary tissue. Parameters were assumed to be probabilistically independent, allowing for efficient parameter ensemble sampling. In an idealized coronary artery geometry, a forward FE model for each parameter ensemble was created to predict tissue stresses under physiologic loading. An emulator was constructed within the UncertainSCI software using polynomial chaos techniques, and statistics and sensitivities were directly computed. Results demonstrated that material parameter uncertainty propagates to variability in predicted stresses across the vessel wall, with the largest dispersions in stress within the adventitial layer. Variability in stress was most sensitive to uncertainties in the anisotropic component of the strain energy function. Moreover, unary and binary interactions within the adventitial layer were the main contributors to stress variance, and the leading factor in stress variability was uncertainty in the stress-like material parameter that describes the contribution of the embedded fibers to the overall artery stiffness. Results from a patient-specific coronary model confirmed many of these findings. Collectively, these data highlight the impact of material property variation on uncertainty in predicted artery stresses and present a pipeline to explore and characterize forward model uncertainty in computational biomechanics.

# 1 Introduction

Physics-based simulations of the cardiovascular system are increasingly being integrated into clinical decision-making (Douglas et al. 2015; Driessen et al. 2019), surgical planning (Trusty et al. 2019), and medical device design (Timmins et al. 2011). Moreover, the U.S. Food and Drug Administration (FDA) has published widely on using simulations to promote the safety, effectiveness, and security of FDA-regulated products (Morrison et al. 2017, 2018; Pathmanathan et al. 2017; Ahmed et al. 2023; Food and Drug Administration: Center for Devices and Radiological Health 2023). As simulations contribute to clinical workflow and regulatory approval and may affect downstream outcomes (e.g., major adverse events, patient death), there is a pressing need to provide confidence in simulation predictions and demonstrate that results are reliable and safe before clinical adoption. Such confidence in simulation pipelines is available in idealized scenarios but is marred by *uncertainties*, which manifest through variability in the subject and clinical variabilities (i.e., model input parameters) that cloud the predictive and prognostic lenses of computer-based modeling. Central to the clinical adoption of patient-specific modeling strategies, therefore, is clearly demonstrating that simulation results are reliable and safe. That is to say, it is essential that simulation results be accompanied by levels of confidence when they potentially impact life-altering decisions.

Advances in medical imaging, computational mechanics, biomechanics, and computing power now enable simulations that predict arterial tissue deformations at the patient-specific level (Taylor and Figueroa 2009; Taylor and Humphrey 2009). In addition to the model geometry, boundary conditions, and numerical approaches to solve the governing equations, the constitutive relation(s) describing the behavior of the material(s) under conditions of interest are required to compute the transmural wall stresses that influence the homeostatic and maladaptive mechanobiological processes. While experimental approaches have been developed to characterize the non-linear, pseudoelastic, and anisotropic material response of vascular tissue under loading, there is much variability in the employed techniques. For example, such methods as ring tests, in-plane biaxial tests, pressure-diameter tests at the *in vivo* length, and biaxial tests consisting of cyclic pressure-diameter and axial force-length protocols have been utilized to characterize the mechanical properties of vascular tissue (see the comprehensive review by Feruzzi *et al.* and references within (Ferruzzi et al. 2013)). Regardless of the specific form of the strain energy function (SEF, $W$), regression analysis can identify the best-fit values of the material parameters within. In addition, there is variability in the applied regression method (e.g., Marquardt-Levenberg) and candidate objective function that is minimized (Humphrey 2002; Ferruzzi et al. 2013). Due to the variability within experimental testing protocols and fitting approaches, as well as variability within and across tissue samples, there exist inherent uncertainties in material parameter(s) describing the soft biologic tissue that propagate to the simulation-predicted mechanical environment.

  In the present study, we incorporated advancements in the field of UQ to evaluate the variability in the output of computational simulations of the arterial mechanical environment due to intrinsic uncertainty in material parameter estimation. In contrast to a traditional deterministic simulation where input parameters have a fixed value that results in a single model output, UQ provides a statistically rigorous approach to determine the influence of input parameter uncertainty by examining a distribution of model outputs (Najm 2009). We applied a novel open-source UQ software tool, `UncertainSCI`, which employs polynomial chaos expansion (PCE) to assess sensitivity, to a forward-modeling framework (Narayan et al. 2022). Therefore, the goal of this study was to leverage PCE UQ to examine the impact of uncertainty in tissue material properties on the variability in model outputs, namely the predicted stress under physiology loading. Given the clinical significance of coronary artery disease and the role of mechanics in the development and progression of the disease (Brown et al. 2016; Tsao et al. 2022), we focused on uncertainty in material characterization and computational models of this vascular territory. To demonstrate the approach, we evaluated models of a generalized multi-layered, thick-walled vessel representative of a coronary artery and a patient-specific model of an epicardial coronary artery.

**2 Methods**

An overview of the integrated UQ-finite element (FE) modeling framework is presented in Fig. 1. Briefly, probability distributions were fit to $n$-material parameters, which were derived from material testing of human coronary tissue, in a structurally-motivated SEF. The parameter space was sampled to generate $m$-parameter ensembles. Utilizing a batch-processing framework, forward FE models for each parameter ensemble were created, and FE analysis was carried out to predict tissue deformation, strains, and stresses. Finally, statistics of the model outputs were computed, and uncertainty was quantified due to material parameter variability.

**Figure 1 here.**

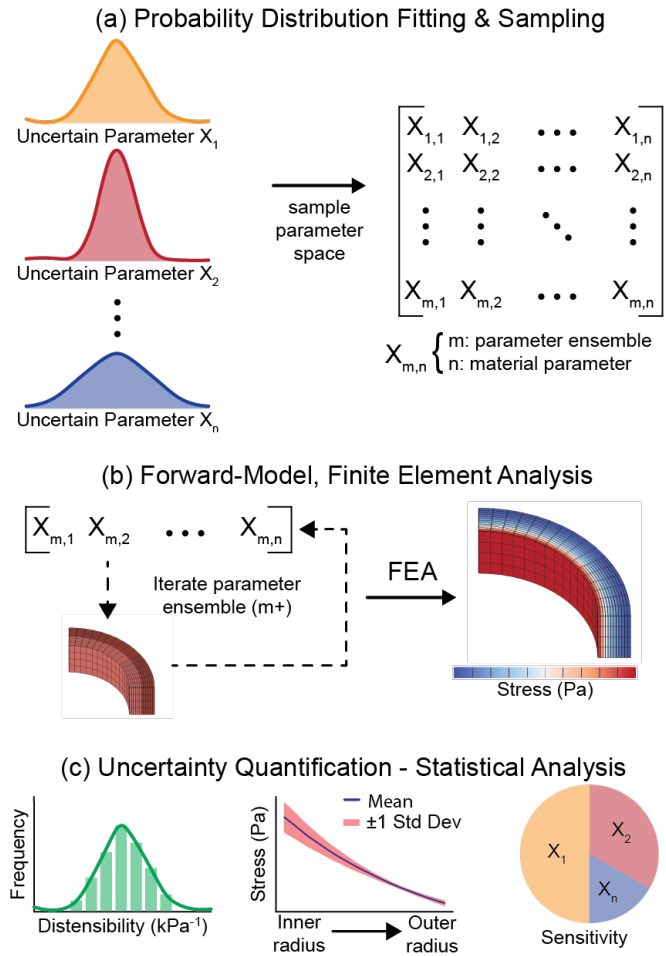

**Fig. 1** Schematic of the integrated uncertainty quantification-finite element analysis (FEA) computational framework. **a** Probability distributions were fit to data on the material parameters ($n$) and `UncertainSCI` was utilized to sample and generate $m$-parameter ensembles. **b** An automated modeling framework assigned each material parameter ensemble to a common `FEBio` input file model and FEA was carried out for $m$-number of models. **c** FE results were post-processed to extract relevant uncertainty quantification metrics and perform statistical analysis.

*2.1. Material Parameter Probability Distribution Sampling*

Material parameters derived from previous layer-specific mechanical behavior testing of 13 human nonatherosclerotic left anterior descending coronary artery tissues were employed herein (Holzapfel 05). Briefly, uniaxial extension testing was performed in the circumferential and longitudinal (axial) directions for the media and adventitia layers. Best-fit material parameters were determined from the mechanical behavior curves using the structurally-motivated SEF,

$$W(\boldsymbol{C}, \boldsymbol{M}^i) = \frac{\mu}{2}(I_1 - 3) + \sum_{i=1,2} \frac{k_1}{2k_2} e^{\{k_2[(1-\rho)(I_1-3)^2 + \rho(I_4^i-1)^2]\}} - 1, \qquad \text{(Eq. 1)}$$

where $\mu$ represents the ground matrix stiffness, $k_1$ is a fiber stress-like parameter, $k_2$ is a dimensionless parameter, $\rho$ is a measure of fiber dispersion within the bounds of [0,1] (0 = no fiber alignment, 1 = perfect fiber alignment along the prescribed vector defined by angle $\phi$), and $I_1$ and $I_4^i$ are the first and fourth invariant, respectively, of the right Cauchy-Green tensor ($\boldsymbol{C}$), defined as,

$$I_1 = \lambda_r^2 + \lambda_\theta^2 + \lambda_z^2, \quad I_4^i = \lambda_\theta^2 \cos^2\phi + \lambda_z^2 \sin^2\phi, \qquad \text{(Eq. 2)}$$

for a deformation gradient that takes the form, $\boldsymbol{F} = diag(\lambda_r, \lambda_\theta, \lambda_z)$. <u>$\boldsymbol{M}^i$ is a unit vector, $(0, \cos\phi^i, \sin\phi^i)$, indicating the orientation of a fiber family, where $\phi^i$ defines the angle between the embedded fiber family and the circumferential axis in the circumferential-axial plane. Two fiber families were considered ($\phi^1 = -\phi^2$), together describing</u>

symmetric fiber families with the same material properties around the long axis of the vessel.

Probability distributions were created for each constitutive parameter in the medial and adventitial layers (10 in total, Fig. 2). Note that parameter distributions were assumed independent, which was guided by the original description of the SEF (Eq. 1) and a lack of data demonstrating any physical relationship amongst them (Holzapfel et al. 2000). Material parameters $\mu$, $k_1$, $k_2$, and $\phi$ employed gamma distributions to avoid non-physiologic parameters (i.e., values had to be >0), whereas $\rho$ used a beta distribution to take advantage of its inherent bounds [0,1]. Probability density functions (PDFs) were fit to the experimental data of each individual parameter using maximum likelihood estimation via the *mle* function in `Matlab`. The material parameter PDFs were defined as distribution objects within `UncertainSCI`, and a parameter ensemble was created from the 10 PDFs that sampled the entire parameter space and accounted for potential output dependence on parameter ensembles (Fig. 1a).

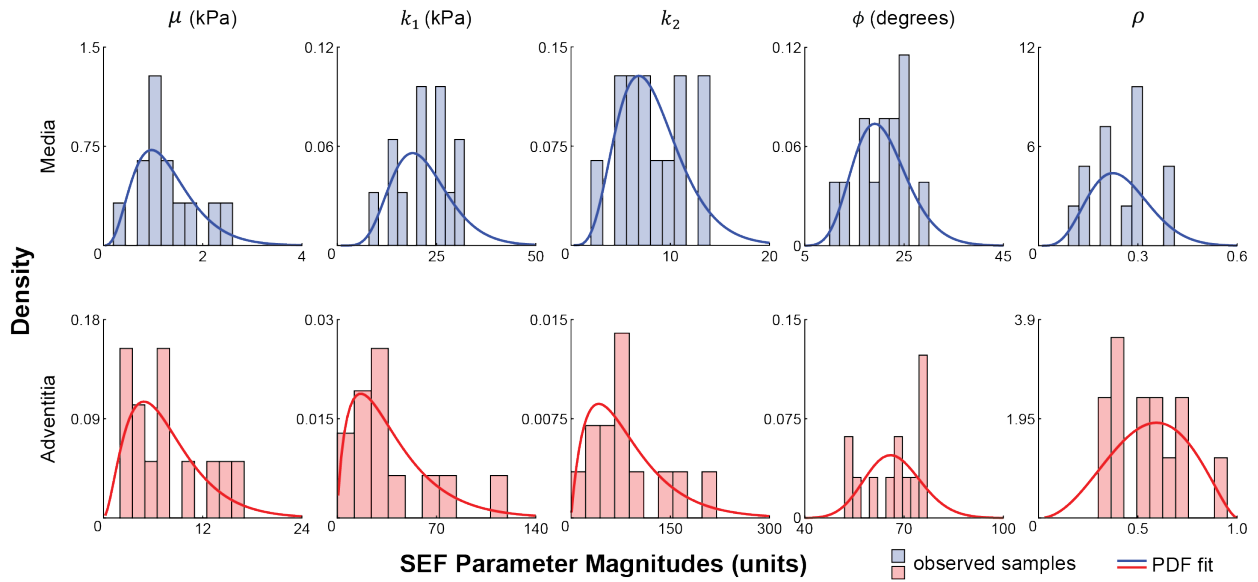

**Fig. 2** Layer-specific material parameter distributions and probability density functions (PDFs). Material parameters from all samples for the media (top row) and adventitia (bottom row) layers from prior uniaxial testing of coronary arteries (Holzapfel et al. 2005). PDFs for each parameter were fit to the observed data. Gamma distributions were fit to data for parameters $\mu$, $k_1$, $k_2$, and $\phi$, and a beta distribution was fit to data for $\rho$.

## 2.2. Idealized Artery Computational Model

A generalized computational model of a human left anterior descending coronary artery was constructed. The artery was modeled as a multi-layered, axisymmetric quarter-cylinder ($L$ = 1 mm, $R_i$ = 1.59 mm, $R_o$ = 2.25 mm), with a medial and adventitial layer thickness of 0.32 and 0.34 mm, respectively (Holzapfel et al. 2005). The intima layer was neglected in this model, as it provides negligible structural support (Burton 1954). The structurally-motivated SEF given by Equation 1, an available material model in `FEBio` (HGO-coronary), described both the medial and adventitial layers. The artery was discretized with 8-node hexahedral "brick" elements with 6 elements in the radial direction for each layer. A mesh convergence study demonstrated that 12 elements in the radial direction (6 elements in each layer) were required to achieve convergence, as a higher mesh density (24 radial elements) led to a ≤2.5% change in the 2-norm criteria for the 1st principal stress (Supp. Fig. 1). The media and adventitia arterial layers were "welded" with shared nodes at the interface. Applied loads included lumen pressures of 80 (diastolic) and 120 (systolic) mmHg. The boundary conditions comprised of fixing the vessel ends in the axial direction and symmetry in the $\theta$-planes to restrict rigid body motions. Quasi-static finite element analysis was performed using the open-source, nonlinear finite element software suite `FEBio` (Maas et al. 2012, 2017). Solver details include using the implicit solver, auto-time stepper (initial and maximum time-step size = 0.1), non-symmetric form of the stiffness matrix, quasi-Newton method (Broyden-Fletcher-G-S; global stiffness matrix reformed each time step), and `PARDISO` linear solver. Solver settings ensured numerical robustness and the ability to support parallel execution. Simulation results were post-processed to evaluate the deformed inner ($r_i$) and outer ($r_o$) radii, transmural distributions of 1st and 3rd principal stresses ($\sigma_1, \sigma_3$), and distensibility ($D$, Eq. 2) (Ferruzzi et al. 2013), which was defined as,

$$D = \frac{d_{i,sys}^2 - d_{i,dias}^2}{d_{i,dias}^2 (P_{sys} - P_{dias})}, \qquad \text{(Eq. 2)}$$

where $d_{i,sys}$ and $d_{i,dias}$ are the deformed inner diameters at systole and diastole, respectively, and $P_{sys}$ and $P_{dias}$ are systolic and diastolic pressure, respectively.

A batch-processing scheme was developed to iterate through material parameter ensembles while using the same `FEBio` input file, which contained information on model mesh and connectivity, material SEFs and parameters, and boundary conditions. A `Matlab` sub-routine was written that iterated through the size-$M$ parameter ensemble, writing $M$ unique input files (Fig. 1b). `FEBio` was called and executed using the `GIBBON` toolbox (M Moerman 2018), and simulation results were stored for UQ analysis. For the idealized artery models, the batch-processing scheme and FE models were run on a Windows 10 server machine with an Intel® Xeon® Silver 4110 CPU (8 cores at 2.10 - 3.00 GHz).

*2.3. Patient-specific Coronary Artery Computational Model*

A three-dimensional representative patient-specific model of the left main and left anterior descending coronary arteries was constructed by expanding established techniques (Samady et al. 2011; Timmins et al. 2015). The end-diastolic geometry was created by fusing bi-plane angiographic image data and virtual-histology intravascular ultrasound (VH-IVUS) images (Fig. 3). Lumen and media-adventitial boundary contours were stacked perpendicular to the IVUS catheter centerline, and catheter torsion was accounted for via the sequential triangular algorithm (Wahle et al. 1999). A medial layer was constructed from smoothed IVUS boundary contours, and an adventitial layer was added with a constant thickness of 400 µm (Waller et al. 1992). Branches were added from IVUS and angiographic-defined locations with branch layer thicknesses derived from post-mortem coronary mean lumen diameter and layer thickness values (Waller et al. 1992). The geometry was meshed with nonlinear tetrahedral elements via `tetGen` and `Gibbon` (Si 2015; Maas et al. 2016; M Moerman 2018), with unique material properties for each layer prescribed using structurally-motivated SEF (Eq. 1) (Holzapfel et al. 2005). To aid the application of boundary conditions, a rectangular box of perivascular (PV) tissue with compressible, neo-Hookean properties ($E$ = 1 kPa, $v$ = 0.3) was added around the coronary geometry and shared identical nodes with the outer vessel surface. The PV outer boundary surfaces were at least 10 mm away from all nodes in the artery (Fig. 3B) and were fixed in all global directions. <u>Preliminary studies on idealized and patient-specific coronary geometries demonstrated that a PV support with those material properties ($E$ = 1 kPa, $v$ = 0.3) and a thickness of 10 mm had a negligible effect on the deformation of the arterial tissue under an applied lumen pressure.</u> Axial motion was prohibited at the vessel and branch end surfaces. The lumen was pressurized to 40 mmHg <u>(note: the reference geometry represented the vessel at end-diastole; ~80 mmHg)</u>. A set of patient-specific models was created using an identical set of material parameter ensembles from the ideal quarter cylinder model at PCE order 3 (n=592 parameter samples). <u>Given the increased complexity of the patient-specific models compared to idealized models, solver and solution parameters were modified. Broyden's quasi-Newton method was employed as the solver. The auto-time stepper parameters were reduced (initial time-step size = 0.01, max time-step size = 0.05), and the *aggressiveness* parameter was turned on to aid the identification of time-step size after a failed step. To aid computational efficiency and solution convergence, the discretization was refined until mesh quality, as determined by the radius-edge ratio in the *Mesh Inspector* feature in FEBio, was deemed suitable. Across the >169k quadratic tetrahedral elements in the patient-specific artery model, excluding</u>

perivascular support, the average radius-edge ratio was 0.934, and <0.003% of the elements had a value >2.5. The batch scheme was executed on two high-performance compute servers (192 Intel Xeon Platinum 8360H CPU @ 3.00GHz cores (HT) per machine) with 14 concurrent jobs running on each machine using 8 cores per simulation (Scientific Computing and Imaging Institute, University of Utah).

**Figure 3 here.**

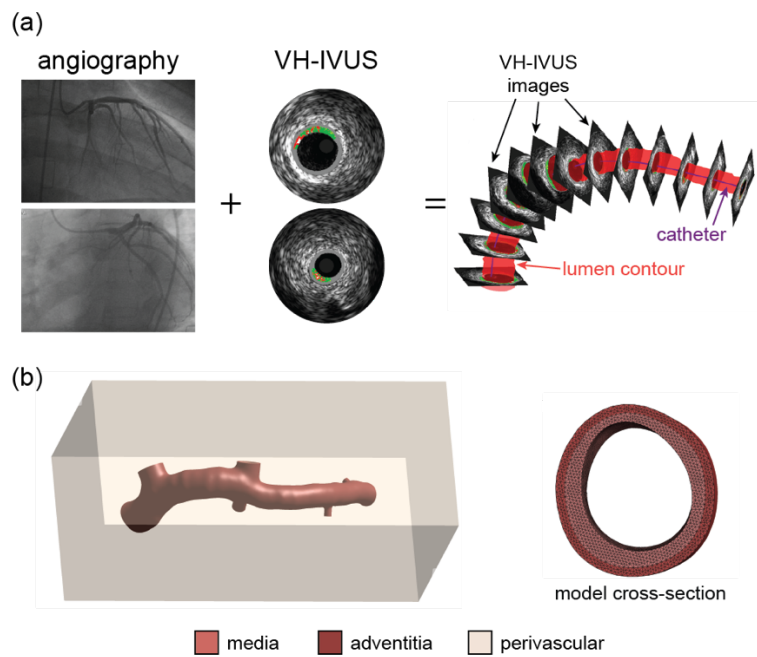

**Fig. 3** Patient-specific coronary model construction. **a** Angiographic and VH-IVUS data were fused to create the 3D lumen and medial-adventitial boundaries. **b** The multi-layer coronary geometry, meshed with tet10 elements, was surrounded by compressible perivascular tissue.

*2.4. Uncertainty Quantification and Sensitivity Analysis*

The open-source, Python-based software suite, `UncertainSCI`, was employed to perform forward model UQ analysis (Narayan et al. 2022). `UncertainSCI` utilizes non-intrusive PCE techniques to query forward model data output over a parameter ensemble (i.e., generated parameter samples) to construct a parameter-to-model-output emulator. With a tunable parameter $p$, the PCE order, the emulator is comprised of a sum of polynomial functions of degree at most $p$, and serves as a surrogate model that approximates the mapping between the input parameter(s) and model output without the need to solve for a computationally expensive forward model. In this study, for example, the constructed emulator provides a relationship between SEF parameters (inputs) and FE-predicted principal strains and stresses (outputs) across the queried range of parameter distributions. `UncertainSCI` constructs parameter samples via the weighted approximate Fekete points method (Guo et al. 2018). The construction of the emulator allows direct extraction of statistics, uncertainty characteristics, and model sensitivity. For a comprehensive description of polynomial chaos techniques and their application, the interested reader is directed to the work by Najm (Najm 2009).

The quality of the PCE was assessed in the idealized computational models by quantifying the *relative error* ($\varepsilon_\delta$) between the PCE approximations and FE-predicted model outputs (e.g., $\sigma_1$), whereby,

$$\varepsilon_\delta(p) = \frac{1}{N}\sum_{i=1}^{N} \frac{\left\|A\hat{x}-\vec{b}\right\|_2}{\left\|\vec{b}\right\|_2}, \qquad \text{(Eq. 3)}$$

where $A\hat{x}$ approximates the solution to the model output $(\vec{b})$, $N$ is the number of elements through the vessel thickness (i.e., transmurally), and $||\cdot||_2$ indicates the 2-norm of the vector. For orders $p = \{1,2,\ldots,5\}$, the relative error was calculated for 5 independent PCE runs. In addition, parameter ensembles were oversampled to evaluate error stability across sampling rates and ensure the aliasing error is minimized[1]. Statistical measures (e.g., mean, standard deviation, coefficient of variation) were calculated directly from the PCE model output. Sensitivity indices, which measure the relative contribution of individual parameters and parameter ensembles to the overall variability of the emulator (i.e., variability in model output), were calculated across the SEF parameters. More specifically, Sobol indices (Sobol′ 2001) were determined to measure the direct effect of an individual parameter (unary interaction; first-order Sobol indices) and parameter ensembles (binary, tertiary, etc. interactions) have on the variance in the model output for 1st principal stress ($\sigma_1$).

**3 Results**

*3.1. PCE Construction, Quality, and Order Convergence*

The number of parameter ensembles generated (i.e., $M$, Fig. 1b), time required to generate these ensembles within `UncertainSCI`, and run time for the FE-batch processing within `FEBio` for the idealized geometry across PCE orders are presented in Table 1. The number of parameter ensembles ranged from ~40 (order 1) to >6,000 (order 5) and required between several minutes to many hours to generate the ensembles and run the FE simulations. Evaluation of the relative error for the 1st and 3rd principal stresses across orders demonstrated reduced error and error variability across multiple PCE runs as order number increased (Fig. 4a,b). Data indicated that order 3 captured the dominant uncertainty modes, as relative error values were <0.4% with a standard deviation of <0.1 across 5 runs. Furthermore, Sobol indices maintained stability at order 3 and higher orders. First-order Sobol indices in the adventitia changed by ≤0.003 and the relative positions remained unchanged across orders 3 to 4 (Fig. 4c). Smaller changes were observed at higher orders (Supp. Fig. 2). A similar trend in the stabilization of the first-order Sobol indices was seen in the medial layer (Supp. Fig. 2). Moreover, second-order Sobol indices and their relative positions were preserved across orders 3 to 5 (Supp. Fig. 3). Examining the FE results from the order 3 simulations demonstrated that a range of deformed geometries and distensibilities were captured (Fig. 5). Across the ~600 parameter ensembles at order 3, deformed inner diameter and thickness values ranged from 3.76-4.78 mm and 0.49-0.58 mm, respectively, and distensibility values ranged from 4.67 to 18.72 MPa$^{-1}$.

---

[1] Pilot studies demonstrated that 2x oversampling notably reduced fluctuations in the Sobol indices and was thus sufficient to ensure stability in PCE results and conclusions drawn (Supp. Fig. 2).

**Table 1 here.**

| Order ($p$) | Parameter Ensembles Generated ($M$) | Parameter Generation Run Time (h:mm:ss) | FE Simulations Run Time (h:mm:ss) |
|---|---|---|---|
| 1 | 42 | 0:01:46 ± 0:00:03 | 0:02:50 ± 0:00:01 |
| 2 | 152 | 0:04:07 ± 0:00:03 | 0:09:26 ± 0:00:03 |
| 3 | 592 | 0:13:45 ± 0:00:06 | 0:36:20 ± 0:00:17 |
| 4 | 2022 | 0:45:12 ± 0:00:06 | 2:03:12 ± 0:00:18 |
| 5 | 6026 | 2:24:14 ± 0:01:48 | 6:10:34 ± 0:03:29 |

**Table 1** Computational times across PCE orders for the idealized artery model. Five PCE runs were performed across each order. Data are reported as mean ± standard deviation.

**Figure 4 here.**

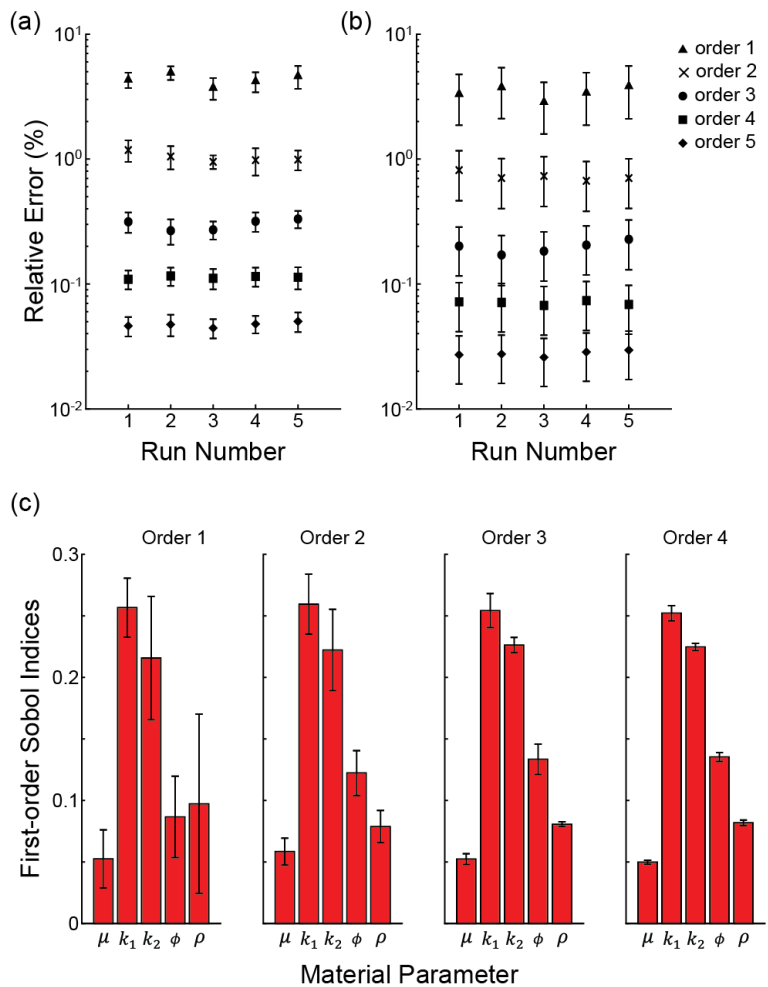

**Fig. 4** PCE and sensitivity indices convergence. Relative error ($\varepsilon_\delta$) in **a** 1$^{st}$ principal stress and **b** 3$^{rd}$ principle stress for PCE analysis across orders and runs. **c** First-order Sobol indices in the adventitia across orders 1 to 4. Data are reported as mean ±

standard deviation. See Supp. Fig. 2 for additional data for order 5 and Sobol indices for the medial layer.

**Figure 5 here.**

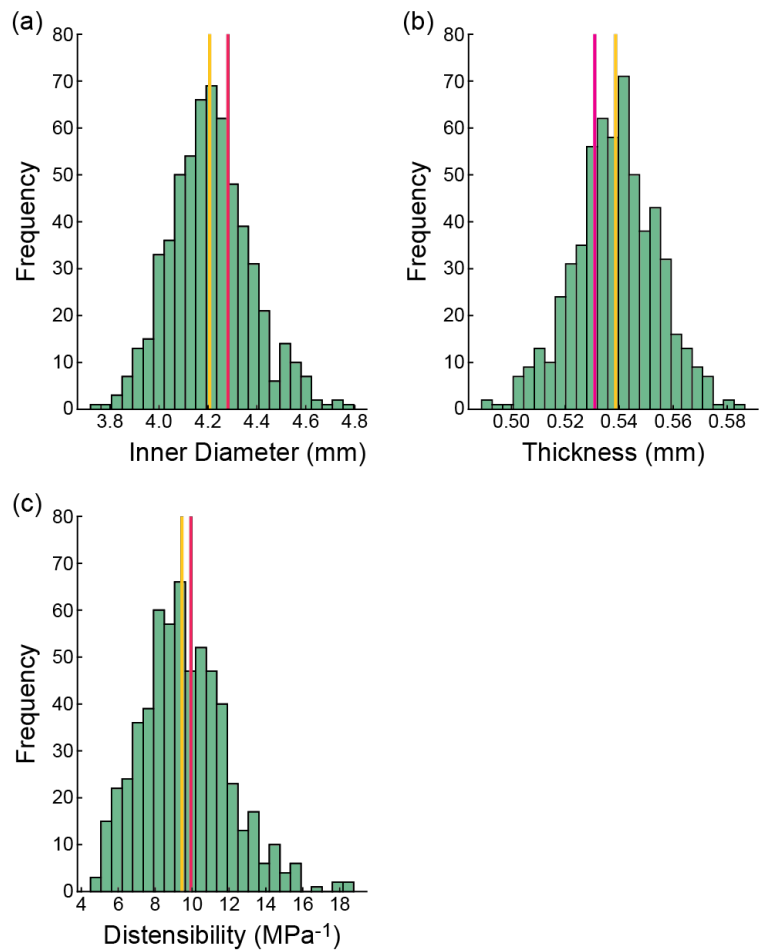

**Fig. 5** Distribution of deformed geometries and structural stiffness from FE models for order 3 PCE analysis. **a** Deformed inner diameter, **b** deformed thickness, and **c** distensibility. Yellow line: median values from UQ-FE models (i.e., median of the output), magenta line: FE model output at median parameter values (i.e., output at the median).

*3.2. UQ and Sensitivity Analysis in Idealized Coronary Model*

The propagation of material parameter uncertainties to the transmural distribution of 1st principal stress at 120 mmHg yielded large deviations from the median within the medial and adventitial layers (Fig. 6a). While median stress values and stress variance decreased radially through each layer, there was an abrupt increase in the variance at the innermost region of the adventitia. Also, variances were higher overall in the adventitia. As a result, coefficient of variation values in the adventitia were >1.5× the values in the media, indicating adventitial stress values had greater dispersion around the mean (Fig. 6b). Sensitivity analysis highlighted that the material parameters in the anisotropic component of the adventitia dominated the variance in the predicted 1st principal stress (Fig. 7). For example, adventitial material parameters $k_1$, $k_2$, $\phi$, and $\rho$ accounted for nearly 70% of the variance in predicted stress values due to a single parameter (unary interactions), with $k_1$ alone accounting for ~25% (i.e., variance in FE-predicted stresses are largely explained by the uncertainty in the stress-like parameter describing the contribution of the adventitial fibers to the artery stiffness, $k_1$). Notably, the uncertainty in the stiffness of the Neo-Hookean ground matrix, controlled by parameter $\mu$, had a negligible effect on stress variance in the media, and only a marginal effect in the adventitia.

**Figure 6 here.**

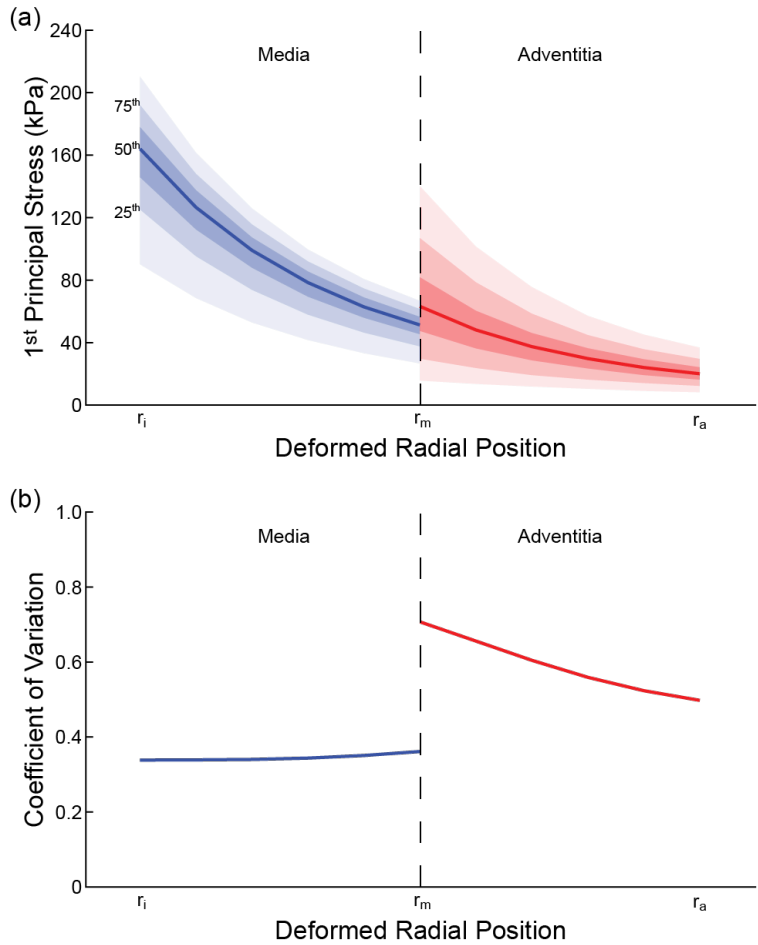

**Fig. 6** Transmural distributions of **a** 1st principal stress (solid line: median, shaded regions: percentile bands) and **b** coefficient of variation from order 3 PCE analysis.

**Figure 7 here.**

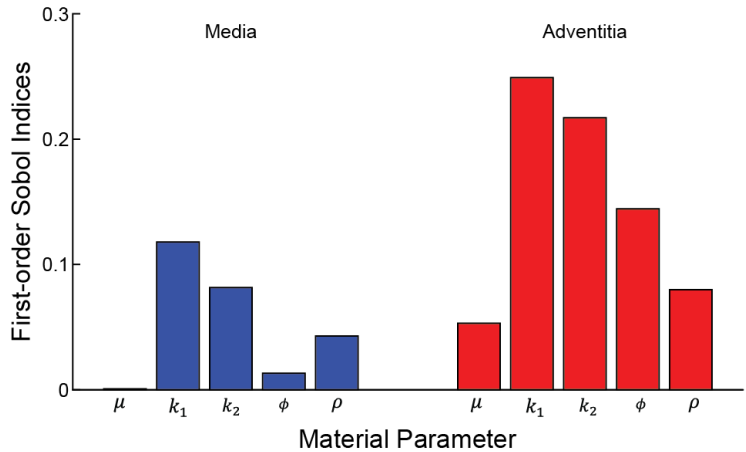

**Fig. 7** Normalized first-order Sobol indices (unary interactions) for material parameters in the media and adventitia. Data were derived from order 3 PCE analysis.

Sensitivity analysis further revealed unique interactions between two parameters (i.e., binary interactions) that contributed to the variance in predicted stress. While unary interactions were dominant, binary interactions still accounted for 12.2% of the variance in predicted 1st principal stress (Fig. 8a). Examining pairwise interactions within each arterial layer highlighted that such interactions in the adventitia accounted for far greater stress variance than those in the media (Fig. 8b). Of the 12.2% of the variance in stress due to binary interactions, the interaction involving the adventitia alone accounted for 46.8% of the variance, and the media-only interaction accounted for 11.6%. The binary interaction of parameters across layers (i.e., inter-layer) accounted for 41.6% of that variance. That is to say, 5.7% of the (total) variance in predicted stress was due to binary interactions between adventitial parameters alone (46.8% of 12.2%), compared to 1.4% for medial parameters and 5.1% for inter-layer parameter combinations. Across the 45 possible pairwise combinations, interactions between the adventitia $\phi$-$\rho$ and $k_2$-$\phi$ dominated, with normalized second-order Sobol indices of 0.21 and 0.10, respectively

(Fig. 8c). Moreover, 5 of the top 10 binary interactions were between adventitial SEF parameters within the anisotropic component. Modest interactions between parameters within the media and adventitia were observed (e.g., media $k_1$, adventitia $k_2$: 0.07), and only 1 binary interaction between medial parameters was in the top 10 (media $k_1$, $k_2$: 0.07). Lastly, tertiary interactions (i.e., interactions between 3 material parameters) accounted for ≤3.1% of the variance in predicted 1st principal stress (Fig. 8a).

**Figure 8 here.**

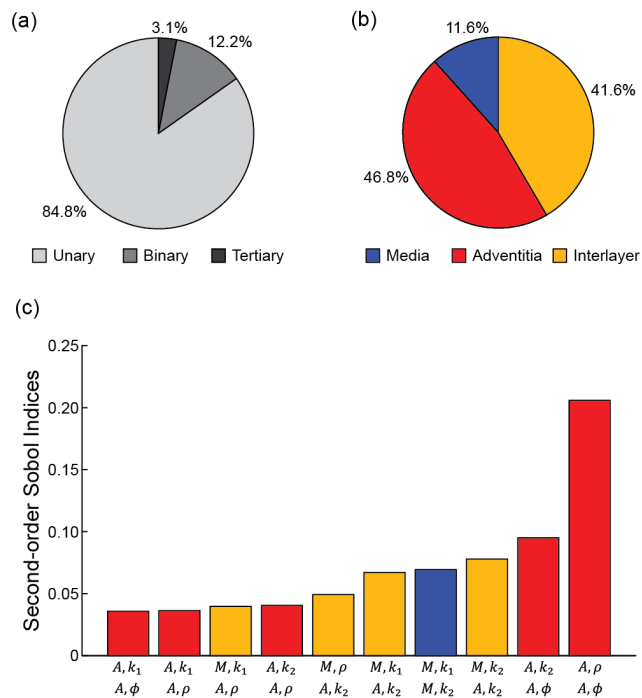

**Fig. 8** Second-order sensitivity analysis from order 3 PCE analysis. **a** Percent output variance due to unary (single parameter), binary (pairwise combinations), and tertiary (triplewise combinations) interactions. **b** Percent output variance due to binary interactions in the media, adventitia, and across layers (i.e., stress variances for the subset attributable to binary interactions). **c** Top ranked second-order Sobol indices

(binary interactions) normalized across combinations. Data were derived from order 3 PCE analysis.

*3.3. Application of UQ to Patient-specific Model of Coronary Artery*

The batch-processing scheme (592 patient-specific models, PCE order 3) completed in ~49.5 hours on the multicore compute servers. Examining the uncertainty in 1st principal stress revealed spatial heterogeneity in statistical and sensitivity measures across the physical domain (Fig. 9a). At a cross section distal from the left circumflex (slice 1; Fig. 9b), for example, mean stresses across the simulations ranged from 15.9-44.6 and 11.8-27.0 kPa in the medial and adventitial layers, respectively. Similar trends of higher values in the media were observed at other spatial locations (Fig. 9b) and when comparing standard deviation and coefficient of variation values across the models. Like the idealized model results, first-order Sobol indices in the patient-specific model associated with adventitial material parameters dominated variances in predicted stress (Fig. 9c). The isotropic and anisotropic material parameters in the adventitia accounted for 23.3% and 42.5% of the variance in stress, respectively. Notably, first-order Sobol indices have marked spatial variation, with increased dispersion in the adventitial layer (Fig. 9c).

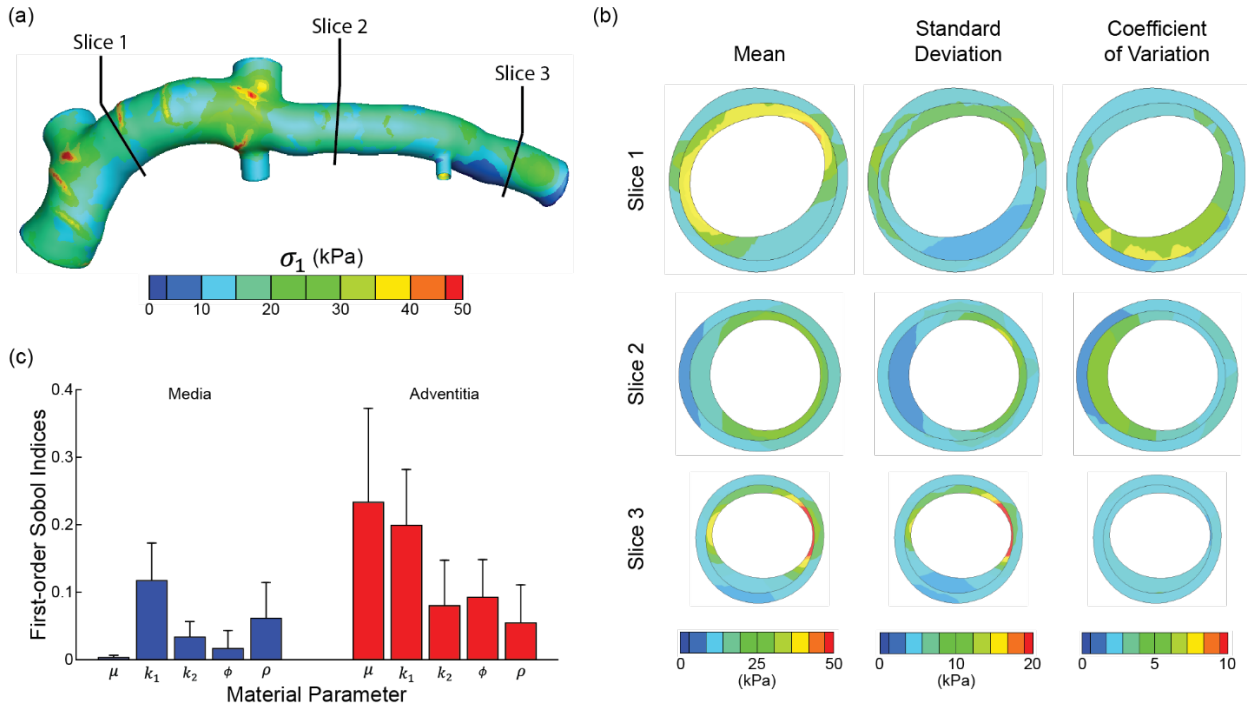

**Fig. 9** Uncertainty quantification and sensitivity analysis in a patient-specific coronary artery model from order 3 PCE analysis (592 simulations). **a** Mean first principal stress ($\sigma_1$) plotted on the unloaded geometry. **b** Transmural distributions of the stress mean, standard deviation, and coefficient of variation plotted in the media and adventitial layers at locations along the coronary vessel. **c** Normalized first-order Sobol indices for material parameters in the media and adventitia (spatially averaged). Data are reported as mean $\pm$ standard deviation across all elements within each material layer.

## 4 Discussion

Herein, we demonstrate the utility of computational UQ in characterizing and quantifying how uncertainties in arterial material parameters propagate to variability in FE-predicted stresses under physiologic loads. In an idealized (cylindrical artery) and patient-specific coronary artery computational model, we show that the variability in predicted stresses is sensitive to uncertainties in the anisotropic component of the material SEF but not uncertainties in the isotropic component. Moreover, unary and binary interactions within the adventitial layer are the main contributors to variance in transmural stresses, with variability in the stiffness of the embedded fibers (i.e., $k_1$) being the leading factor in stress variability. Lastly, we highlight the non-intrusive nature of `UncertainSCI` and the ability to couple this powerful yet lightweight UQ framework to the `FEBio` software suite.

Given that vascular tissue has an anisotropic structural organization and a non-linear stress-strain response, it was not surprising to see that uncertainty in the anisotropic SEF parameters accounted for the greatest variance in stress (Figs. 7,8). Moreover, sensitivity analysis demonstrated that the unary interaction of the initial stiffness of the fibers ($k_1$) and the binary interaction between fiber angle ($\phi$) and fiber dispersion ($\rho$) in the adventitia layer were prominent parameters that influence stress variabilities. Experimental data highlight that the adventitia is stiffer than the media layer (Holzapfel 2005), resulting from the dense network of type I collagen within the ground matrix. Thus, the presented results provide further evidence of the influence of the arterial structure on the mechanical behavior of this soft tissue. Also, it is important to recognize the greater dispersion of the anisotropic SEF parameters in the adventitia versus media (Fig. 2), contributing to the larger coefficient of variation and Sobol indices in the adventitia. In addition to increased sample testing, advances in experimental approaches to better characterize and quantify these SEF parameters will promote reduced uncertainty in calculated stress values.

Efforts have been focused on integrating computational models of the cardiovascular system into the clinical setting; however, multiple sources of uncertainty must be accounted for to provide confidence in model predictions. In patient-specific models, for example, uncertainty arises in domain (geometry) construction, boundary conditions, numerical schemes, and, as investigated herein, material properties. In the context of material properties, sources of uncertainty are present when utilizing either population-based or patient-specific data. A population-based approach was used in this study, whereby variability across patient samples contributed to the uncertainty in arterial stiffness (Holzapfel et al. 2005). Importantly, this approach requires quantifying arterial mechanical properties on large data sets utilizing standardized protocols, including biaxial material characterization, to minimize experimental variance (Walsh et al. 2014). While methods to non-invasively determine patient-specific material properties are in development (e.g., DENSE MRI (Bracamonte et al. 2020)), there remain limitations with such strategies that must be resolved before adoption. Moreover, uncertainty exists in patient-specific strategies due to material spatial heterogeneity and image noise that must be accounted for within the modeling framework. Despite longitudinal studies that have utilized deterministic modeling approaches to demonstrate the utility of arterial wall stress as a prognostic marker of, for example, coronary plaque or abdominal aneurysm rupture (Teng et al. 2014; Polzer et al. 2020), effective modeling strategies and decision guidelines must be robust to uncertainty and provide levels of reliability and safety before clinical adoption.

Capturing the randomness of the input parameters (i.e., defining accurate PDFs) is central to parametric UQ, whereby the randomness of input parameters propagates forward to the model outputs. The accuracy of constructed PDFs to capture SEF material parameter distributions depends on whether the evaluated samples (observations) sufficiently explore the parameter space. While mechanical testing data on human tissue samples across the vascular tree are reported (Vande Geest et al. 2004; Holzapfel et al. 2005; Teng et al. 2009), these studies are often limited to a few samples, which may not provide sufficient numbers to define representative PDFs. This study fit PDFs to SEF parameters derived from the experimental testing of 13 human coronary tissue samples, which is the largest reported data set on layer-specific mechanical testing of human coronaries (Holzapfel et al. 2005). The fit PDFs capture the distribution of observations (Fig. 2); however, it is unclear if these fits represent the population distribution and whether the selected distributions are the best descriptors. Indeed, a study reported biaxial testing data from 125 human femoropopliteal arteries and provided a comprehensive analysis that examined differences in material properties across age and disease severity (Jadidi et al. 2021). Yet, given the difficulty in procuring human tissues, particularly healthy samples, such studies are rare, <u>especially in the case of coronary arteries</u>. Thus, the standardization of testing protocols is warranted to allow consolidation of data sets towards improved characterization of population distributions and quantification of parametric UQ input parameter randomness.

An advantage of `UncertainSCI` is that it utilizes non-intrusive UQ techniques, which do not require changes to existing simulation frameworks or numerical schemes, to calculate accurate statistical measures of the forward propagation of uncertainty. Moreover, `UncertainSCI` uses PCE techniques, which are more efficient and offer better convergence than Monte-Carlo (MC) and quasi-MC-based approaches (Xiu and

Hesthaven 2005) and offer advantages in biomedical simulations, where the dependence on parameters is often smooth. Although PCE and MC approaches sample the multivariate parameter distributions, MC approaches require more samples to obtain reliable sensitivity measures and are thus more computationally demanding (Eck et al. 2016). <u>Regardless of the sampling approach, and particularly relevant to soft biological tissue, it must be ensured that the sampled parameter ensembles produce physical stress-stretch responses (Robertson and Cook 2014).</u> While MC methods have been successfully applied to cardiovascular simulations (Sankaran and Marsden 2011; Tran et al. 2019), these approaches can be problematic for complex patient-specific models (e.g., Figs. 3 and 9). Alternatively, PCE approaches compute equivalent statistical metrics, with orders of magnitude fewer evaluation samples compared to MC approaches (Eck et al. 2016; Burk et al. 2020). Importantly, however, PCE approaches are only recommended when the number of uncertain parameters is limited, typically less than 20, after which PCE strategies are no longer more efficient than MC methods (Xiu and Hesthaven 2005; Crestaux et al. 2009; Eck et al. 2016). While the presented study evaluated 10 parameters describing the material properties for the UQ analysis (Eq.1, Fig. 2), additional areas of uncertainty are present in cardiovascular simulations, as discussed above, that would increase UQ complexity and computational demand. Methods can be employed to reduce the number of uncertain inputs. For example, if uncertain inputs have minimal effects on model output variance (i.e., small first-order Sobol indices, Fig. 7), those inputs can be fixed within their uncertain domain. Recently, a novel UQ framework that utilized a multilevel multi-fidelity MC estimator, which incorporates results from zero and one-dimensional models across mesh (spatial) resolutions to efficiently construct estimators, was shown to greatly reduce computational costs (10 to 100× reduction) for UQ in hemodynamic simulations (Fleeter et al. 2020). Continued advancements in data-efficient UQ methods to promote clinical translation and adoption are warranted.

The lack of patient-specific material properties brings into question the reliability of existing coronary artery model results. Without directly comparing results derived from patient-specific or population-based material properties, a few remarks can be made regarding reliability. First, advances in medical imaging, hemodynamic assessment, and constitutive modeling have promoted patient-specific modeling capabilities beyond 2D, linear elastic computational models derived from histology data (Cheng et al. 1993). Thus, even with the lack of material property data, patient-specific models better represent the *in vivo* anatomy and physiology, and model results are more reliable. Second, even with material property limitations, current modeling efforts enable hypothesis generation and testing that have revealed new insights into cardiovascular biology and medicine. As an example, early observations and hypotheses on the role of plaque stress in coronary plaque rupture motivated studies that have realized these observations in patients with acute coronary syndrome (Richardson et al. 1989; Teng et al. 2014). Third, the presented UQ analysis indeed provides an assessment of the reliability of modeling results given material property uncertainty. Thus, studies can (and should) address the interpretation of their results within the context of our presented findings. What remains unknown is how the uncertainty in these data translates to correlations with clinical observations or outcomes, promoting the predictive power of biomechanical indices.

The presented results have implications for interpreting correlations between biological processes, such as tissue homeostasis, growth and remodeling, and disease progression, and the mechanical environment, such as stresses and strains (i.e., vascular mechanobiology) (Humphrey and Schwartz 2021). Whether utilizing analytical solutions or FE approaches to determine stresses within thick-walled vascular tissue, variability in calculated/predicted stresses due to material parameter uncertainty could impact the drawn correlations. Moreover, stress mediated growth laws (i.e., constitutive relations that describe cellular and extracellular matrix produce and removal rates as a function of stress) are central to constrained mixture models of soft tissue growth and remodeling (Humphrey 2021). Recognizing and quantifying the impact of uncertainty on these relationships is critical to advance the understanding of the evolution of soft tissue geometry, composition, and material behavior under complex loading.

There are limitations in this study that should be acknowledged; however, these limitations do not detract from the significance of the results. First, not all possible PDFs were explored to describe the distribution of the reported SEF material parameters. While PDFs that yielded non-physical material parameters were excluded, continued investigation of PDFs that best describe the experimental data is warranted. Second, residual strain was not included in the computational models. While residual strain homogenizes the stress field in the artery wall under physiologic loading (Chuong and Fung 1986), these do not mitigate the variance in wall stresses due to uncertainty in material parameters. Importantly, the opening angle, the geometric quantity that aids quantification of the displacement field to determine residual strain, is yet another variable with uncertainty to evaluate and determine its impact on arterial stress field variance. Third, material testing data on the intimal layer reported in the Holzapfel *et al.* study (Holzapfel et al. 2005) was not incorporated into the idealized model. These data were derived from the mechanical testing of specimens with non-atherosclerotic intimal thickening and diffuse intimal hyperplasia, which are detectable with VH-IVUS imaging (García-García et al. 2009). Future investigations utilizing the presented FE-UQ framework will seek to incorporate additional tissue components and plaque phenotypes. Lastly, only one patient-specific geometry was explored. Although the coronary anatomy and spatial variation in sensitivity measures impart complexities that make it difficult to draw immediate conclusions, the demonstrated application of the UQ-FE framework provides a novel approach for future investigations.

## 5 Conclusion

In summary, we present a computational framework to explore, characterize, and quantify forward model uncertainty in FE simulations of the arterial wall biomechanical environment. We report that uncertainties in SEF parameters describing the material response of a multi-layered, thick-walled artery under physiologic loading are pushed forward, leading to considerable variances in transmural stress fields. These data highlight that there remains a pressing need to promote experimental data collection towards better characterizing SEF material parameter distributions, and further understanding the propagation of such uncertainty to the predicted kinematics and stresses. Moreover, our efforts demonstrate the demand for continued rigor in computational biomechanics by providing confidence in calculated stress metrics to address complex biological and clinical problems.

**Statements and Declarations**

**Author contributions.** Conceptualization: R.S.M., A.N., L.H.T.; Methodology: all authors; Formal analysis and investigation: all authors; Writing - original draft preparation: C.C.B., D.J., Y.F.J.W., L.H.T.; Writing - review and editing: all authors; Funding acquisition: R.S.M., A.N., L.H.T.

**Competing Interests.** The authors have no competing financial or non-financial interests to declare that are relevant to the content of this article.

**Funding.** This work was supported, in part, by the National Institutes of Health grants R01 HL-150608 (L.H.T.), U24 EB-029012 (R.S.M., A.N.), P41 GM-103545 (R.S.M.), R24 GM-136986 (R.S.M.), American Heart Association grant 23PRE1019455 (C.C.B.), National Science Foundation GRFP (L.C.R.), and the Nora Eccles Treadwell Foundation for Cardiovascular Research (R.S.M.).

**Tables**

| Order ($p$) | Parameter Ensembles Generated ($M$) | Parameter Generation Run Time (h:mm:ss) | FE Simulations Run Time (h:mm:ss) |
|---|---|---|---|
| 1 | 42 | 0:01:46 ± 0:00:03 | 0:02:50 ± 0:00:01 |
| 2 | 152 | 0:04:07 ± 0:00:03 | 0:09:26 ± 0:00:03 |
| 3 | 592 | 0:13:45 ± 0:00:06 | 0:36:20 ± 0:00:17 |
| 4 | 2022 | 0:45:12 ± 0:00:06 | 2:03:12 ± 0:00:18 |
| 5 | 6026 | 2:24:14 ± 0:01:48 | 6:10:34 ± 0:03:29 |

**Table 1** Computational times across PCE orders for the idealized artery model. Five PCE runs were performed across each order. Data are reported as mean ± standard deviation.

**Figure Captions**

**Fig. 1** Schematic of the integrated uncertainty quantification-finite element analysis (FEA) computational framework. **a** Probability distributions were fit to data on the material parameters ($n$) and `UncertainSCI` was utilized to sample and generate $m$-parameter ensembles. **b** An automated modeling framework assigned each material parameter ensemble to a common `FEBio` input file model and FEA was carried out for $m$-number of models. **c** FE results were post-processed to extract relevant uncertainty quantification metrics and perform statistical analysis.

**Fig. 2** Layer-specific material parameter distributions and probability density functions (PDFs). Material parameters from all samples for the media (top row) and adventitia (bottom row) layers from prior uniaxial testing of coronary arteries (Holzapfel et al. 2005). PDFs for each parameter were fit to the observed data. Gamma distributions were fit to data for parameters $\mu$, $k_1$, $k_2$, and $\phi$, and a beta distribution was fit to data for $\rho$.

**Fig. 3** Patient-specific coronary model construction. **a** Angiographic and VH-IVUS data were fused to create the 3D lumen and medial-adventitial boundaries. **b** The multi-layer coronary geometry, meshed with tet10 elements, was surrounded by compressible perivascular tissue.

**Fig. 4** PCE and sensitivity indices convergence. Relative error ($\varepsilon_\delta$) in **a** 1$^{st}$ principal stress and **b** 3$^{rd}$ principle stress for PCE analysis across orders and runs. **c** First-order

Sobol indices in the adventitia across orders 1 to 4. Data are reported as mean ± standard deviation. See Supp. Fig. 1 for additional data for order 5 and Sobol indices for the medial layer.

**Fig. 5** Distribution of deformed geometries and structural stiffness from FE models for order 3 PCE analysis. **a** Deformed inner diameter, **b** deformed thickness, and **c** distensibility. Yellow line: median values from UQ-FE models (i.e., median of the output), magenta line: FE model output at median parameter values (i.e., output at the median).

**Fig. 6** Transmural distributions of **a** $1^{st}$ principal stress (solid line: median, shaded regions: percentile bands) and **b** coefficient of variation from order 3 PCE analysis.

**Fig. 7** Normalized first-order Sobol indices (unary interactions) for material parameters in the media and adventitia. Data were derived from order 3 PCE analysis.

**Fig. 8** Second-order sensitivity analysis from order 3 PCE analysis. **a** Percent output variance due to unary (single parameter), binary (pairwise combinations), and tertiary (triplewise combinations) interactions. **b** Percent output variance due to binary interactions in the media, adventitia, and across layers (i.e., interlayer). **c** Top ranked second-order Sobol indices (binary interactions) normalized across combinations. Data were derived from order 3 PCE analysis.

**Fig. 9** Uncertainty quantification and sensitivity analysis in a patient-specific coronary artery model from order 3 PCE analysis (592 simulations). **a** Mean first principal stress ($\sigma_1$) plotted on the unloaded geometry. **b** Transmural distributions of the stress mean, standard deviation, and coefficient of variation plotted in the media and adventitial layers at locations along the coronary vessel. **c** Normalized first-order Sobol indices for material parameters in the media and adventitia (spatially averaged). Data are reported as mean ± standard deviation across all elements within each material layer.

# References

Ahmed KBR, Pathmanathan P, Kabadi SV, et al (2023) Editorial on the FDA Report on "Successes and Opportunities in Modeling & Simulation for FDA." Ann Biomed Eng 51:6–9. https://doi.org/10.1007/s10439-022-03112-x

Bracamonte JH, Wilson JS, Soares JS (2020) Assessing Patient-Specific Mechanical Properties of Aortic Wall and Peri-Aortic Structures From In Vivo DENSE Magnetic Resonance Imaging Using an Inverse Finite Element Method and Elastic Foundation Boundary Conditions. J Biomech Eng 142:. https://doi.org/10.1115/1.4047721

Brown AJ, Teng Z, Evans PC, et al (2016) Role of biomechanical forces in the natural history of coronary atherosclerosis. Nat Rev Cardiol 13:210–220. https://doi.org/10.1038/nrcardio.2015.203

Burk KM, Narayan A, Orr JA (2020) Efficient sampling for polynomial chaos-based uncertainty quantification and sensitivity analysis using weighted approximate Fekete points. Int J Numer Method Biomed Eng 36:e3395. https://doi.org/10.1002/cnm.3395

Burton AC (1954) Relation of structure to function of the tissues of the wall of blood vessels. Physiol Rev 34:619–642. https://doi.org/10.1152/physrev.1954.34.4.619

Cheng GC, Loree HM, Kamm RD, et al (1993) Distribution of circumferential stress in ruptured and stable atherosclerotic lesions. A structural analysis with histopathological correlation. Circulation 87:1179–1187. https://doi.org/10.1161/01.cir.87.4.1179

Chuong CJ, Fung YC (1986) On residual stresses in arteries. J Biomech Eng 108:189–192. https://doi.org/10.1115/1.3138600

Crestaux T, Le Maı̂tre O, Martinez J-M (2009) Polynomial chaos expansion for sensitivity analysis. Reliab Eng Syst Saf 94:1161–1172. https://doi.org/10.1016/j.ress.2008.10.008

Douglas PS, Pontone G, Hlatky MA, et al (2015) Clinical outcomes of fractional flow reserve by computed tomographic angiography-guided diagnostic strategies vs. usual care in patients with suspected coronary artery disease: the prospective longitudinal trial of FFR(CT): outcome and resource impacts study. Eur Heart J 36:3359–3367. https://doi.org/10.1093/eurheartj/ehv444

Driessen RS, Danad I, Stuijfzand WJ, et al (2019) Comparison of Coronary Computed Tomography Angiography, Fractional Flow Reserve, and Perfusion Imaging for Ischemia Diagnosis. J Am Coll Cardiol 73:161–173. https://doi.org/10.1016/j.jacc.2018.10.056

Eck VG, Donders WP, Sturdy J, et al (2016) A guide to uncertainty quantification and sensitivity analysis for cardiovascular applications. Int J Numer Method Biomed Eng 32:. https://doi.org/10.1002/cnm.2755

Ferruzzi J, Bersi MR, Humphrey JD (2013) Biomechanical phenotyping of central arteries in health and disease: advantages of and methods for murine models. Ann Biomed Eng 41:1311–1330. https://doi.org/10.1007/s10439-013-0799-1

Fleeter CM, Geraci G, Schiavazzi DE, et al (2020) Multilevel and multifidelity uncertainty

quantification for cardiovascular hemodynamics. Comput Methods Appl Mech Eng 365:. https://doi.org/10.1016/j.cma.2020.113030

Food and Drug Administration: Center for Devices and Radiological Health (2023) FDA Guidance: Assessing the Credibility of Computational Modeling and Simulation in Medical Device Submissions. U.S. Food and Drug Administration

García-García H, Mintz G, Lerman A, et al (2009) Tissue characterisation using intravascular radiofrequency data analysis: recommendations for acquisition, analysis, interpretation and reporting. EuroIntervention 5:177–189. https://doi.org/10.4244/eijv5i2a29 PMID - 20449928

Guo L, Narayan A, Yan L, Zhou T (2018) Weighted Approximate Fekete Points: Sampling for Least-Squares Polynomial Approximation. SIAM J Sci Comput 40:A366–A387. https://doi.org/10.1137/17M1140960

Holzapfel GA, Gasser TC, Ogden RW (2000) A New Constitutive Framework for Arterial Wall Mechanics and a Comparative Study of Material Models. Journal of elasticity and the physical science of solids 61:1–48. https://doi.org/10.1023/a:1010835316564

Holzapfel GA, Sommer G, Gasser CT, Regitnig P (2005) Determination of layer-specific mechanical properties of human coronary arteries with nonatherosclerotic intimal thickening and related constitutive modeling. American Journal of Physiology-Heart and Circulatory Physiology 289:H2048–H2058. https://doi.org/10.1152/ajpheart.00934.2004

Humphrey JD (2021) Constrained Mixture Models of Soft Tissue Growth and Remodeling - Twenty Years After. J Elast 145:49–75. https://doi.org/10.1007/s10659-020-09809-1

Humphrey JD (2002) Cardiovascular Solid Mechanics, 1st edn. Springer-Verlag New York

Humphrey JD, Schwartz MA (2021) Vascular Mechanobiology: Homeostasis, Adaptation, and Disease. Annu Rev Biomed Eng 23:1–27. https://doi.org/10.1146/annurev-bioeng-092419-060810

Jadidi M, Razian SA, Anttila E, et al (2021) Comparison of morphometric, structural, mechanical, and physiologic characteristics of human superficial femoral and popliteal arteries. Acta Biomater 121:431–443. https://doi.org/10.1016/j.actbio.2020.11.025

M Moerman K (2018) GIBBON: The geometry and image-based bioengineering add-on. J Open Source Softw 3:506. https://doi.org/10.21105/joss.00506

Maas SA, Ateshian GA, Weiss JA (2017) FEBio: History and Advances. Annu Rev Biomed Eng 19:279–299. https://doi.org/10.1146/annurev-bioeng-071516-044738 PMID - 28633565

Maas SA, Ellis BJ, Ateshian GA, Weiss JA (2012) FEBio: Finite Elements for Biomechanics. J Biomech Eng 134:011005. https://doi.org/10.1115/1.4005694 PMID - 22482660

Maas SA, Ellis BJ, Rawlins DS, Weiss JA (2016) Finite element simulation of articular contact mechanics with quadratic tetrahedral elements. J Biomech 49:659–667. https://doi.org/10.1016/j.jbiomech.2016.01.024 PMID - 26900037

Morrison TM, Dreher ML, Nagaraja S, et al (2017) The Role of Computational Modeling and Simulation in the Total Product Life Cycle of Peripheral Vascular Devices. J Med Device 11:. https://doi.org/10.1115/1.4035866

Morrison TM, Pathmanathan P, Adwan M, Margerrison E (2018) Advancing Regulatory Science With Computational Modeling for Medical Devices at the FDA's Office of Science and Engineering Laboratories. Front Med 5:241. https://doi.org/10.3389/fmed.2018.00241

Najm HN (2009) Uncertainty Quantification and Polynomial Chaos Techniques in Computational Fluid Dynamics. Annu Rev Fluid Mech 41:35–52. https://doi.org/10.1146/annurev.fluid.010908.165248

Narayan A, Liu Z, Bergquist JA, et al (2022) UncertainSCI: Uncertainty quantification for computational models in biomedicine and bioengineering. Comput Biol Med 106407. https://doi.org/10.1016/j.compbiomed.2022.106407

Pathmanathan P, Gray RA, Romero VJ, Morrison TM (2017) Applicability Analysis of Validation Evidence for Biomedical Computational Models. J Verif Valid Uncert 2:021005. https://doi.org/10.1115/1.4037671

Polzer S, Gasser TC, Vlachovský R, et al (2020) Biomechanical indices are more sensitive than diameter in predicting rupture of asymptomatic abdominal aortic aneurysms. J Vasc Surg 71:617-626.e6. https://doi.org/10.1016/j.jvs.2019.03.051

Richardson PD, Davies MJ, Born GVR (1989) INFLUENCE OF PLAQUE CONFIGURATION AND STRESS DISTRIBUTION ON FISSURING OF CORONARY ATHEROSCLEROTIC PLAQUES. Lancet 334:941–944. https://doi.org/10.1016/s0140-6736(89)90953-7 PMID - 2571862

Robertson D, Cook D (2014) Unrealistic statistics: how average constitutive coefficients can produce non-physical results. J Mech Behav Biomed Mater 40:234–239. https://doi.org/10.1016/j.jmbbm.2014.09.006

Samady H, Eshtehardi P, McDaniel MC, et al (2011) Coronary Artery Wall Shear Stress Is Associated With Progression and Transformation of Atherosclerotic Plaque and Arterial Remodeling in Patients With Coronary Artery Disease. Circulation 124:779–788. https://doi.org/10.1161/circulationaha.111.021824

Sankaran S, Marsden AL (2011) A stochastic collocation method for uncertainty quantification and propagation in cardiovascular simulations. J Biomech Eng 133:031001. https://doi.org/10.1115/1.4003259

Si H (2015) TetGen, a Delaunay-Based Quality Tetrahedral Mesh Generator. ACM Trans Math Softw 41:1–36. https://doi.org/10.1145/2629697

Sobol′ IM (2001) Global sensitivity indices for nonlinear mathematical models and their Monte Carlo estimates. Math Comput Simul 55:271–280. https://doi.org/10.1016/S0378-4754(00)00270-6

Taylor CA, Figueroa CA (2009) Patient-specific modeling of cardiovascular mechanics. Annu Rev Biomed Eng 11:109–134.

https://doi.org/10.1146/annurev.bioeng.10.061807.160521

Taylor CA, Humphrey JD (2009) Open Problems in Computational Vascular Biomechanics: Hemodynamics and Arterial Wall Mechanics. Comput Methods Appl Mech Eng 198:3514–3523. https://doi.org/10.1016/j.cma.2009.02.004

Teng Z, Brown AJ, Calvert PA, et al (2014) Coronary Plaque Structural Stress Is Associated With Plaque Composition and Subtype and Higher in Acute Coronary Syndrome. Circ Cardiovasc Imaging 7:461–470. https://doi.org/10.1161/circimaging.113.001526

Teng Z, Tang D, Zheng J, et al (2009) An experimental study on the ultimate strength of the adventitia and media of human atherosclerotic carotid arteries in circumferential and axial directions. J Biomech 42:2535–2539. https://doi.org/10.1016/j.jbiomech.2009.07.009

Timmins LH, Miller MW, Clubb FJ Jr, Moore JE Jr (2011) Increased artery wall stress post-stenting leads to greater intimal thickening. Lab Invest 91:955–967. https://doi.org/10.1038/labinvest.2011.57

Timmins LH, Molony DS, Eshtehardi P, et al (2015) Focal Association Between Wall Shear Stress and Clinical Coronary Artery Disease Progression. Ann Biomed Eng 43:94–106. https://doi.org/10.1007/s10439-014-1155-9 PMID - 25316593

Tran JS, Schiavazzi DE, Kahn AM, Marsden AL (2019) Uncertainty quantification of simulated biomechanical stimuli in coronary artery bypass grafts. Comput Methods Appl Mech Eng 345:402–428. https://doi.org/10.1016/j.cma.2018.10.024

Trusty PM, Wei ZA, Slesnick TC, et al (2019) The first cohort of prospective Fontan surgical planning patients with follow-up data: How accurate is surgical planning? J Thorac Cardiovasc Surg 157:1146–1155. https://doi.org/10.1016/j.jtcvs.2018.11.102

Tsao CW, Aday AW, Almarzooq ZI, et al (2022) Heart Disease and Stroke Statistics-2022 Update: A Report From the American Heart Association. Circulation 145:e153–e639. https://doi.org/10.1161/CIR.0000000000001052

Vande Geest JP, Sacks MS, Vorp DA (2004) Age dependency of the biaxial biomechanical behavior of human abdominal aorta. J Biomech Eng 126:815–822. https://doi.org/10.1115/1.1824121

Wahle A, Prause GPM, DeJong SC, Sonka M (1999) Geometrically correct 3-D reconstruction of intravascular ultrasound images by fusion with biplane angiography-methods and validation. IEEE Trans Med Imaging 18:686–699. https://doi.org/10.1109/42.796282 PMID - 10534051

Waller BF, Orr CM, Slack JD, et al (1992) Anatomy, histology, and pathology of coronary arteries: a review relevant to new interventional and imaging techniques--Part I. Clin Cardiol 15:451–457. https://doi.org/10.1002/clc.4960150613

Walsh MT, Cunnane EM, Mulvihill JJ, et al (2014) Uniaxial tensile testing approaches for characterisation of atherosclerotic plaques. J Biomech 47:793–804. https://doi.org/10.1016/j.jbiomech.2014.01.017 PMID - 24508324

Xiu D, Hesthaven JS (2005) High-Order Collocation Methods for Differential Equations with Random Inputs. SIAM J Sci Comput 27:1118–1139. https://doi.org/10.1137/040615201

**Supplementary Material**

Author names: Caleb C. Berggren, David Jiang, Y.F. Jack Wang, Jake A. Bergquist, Lindsay C. Rupp, Zexin Liu, Rob S. MacLeod, Akil Narayan, Lucas H. Timmins

Corresponding author: Lucas H. Timmins, Texas A&M University,

lucas.timmins@tamu.edu

**Supplementary Figure 1**

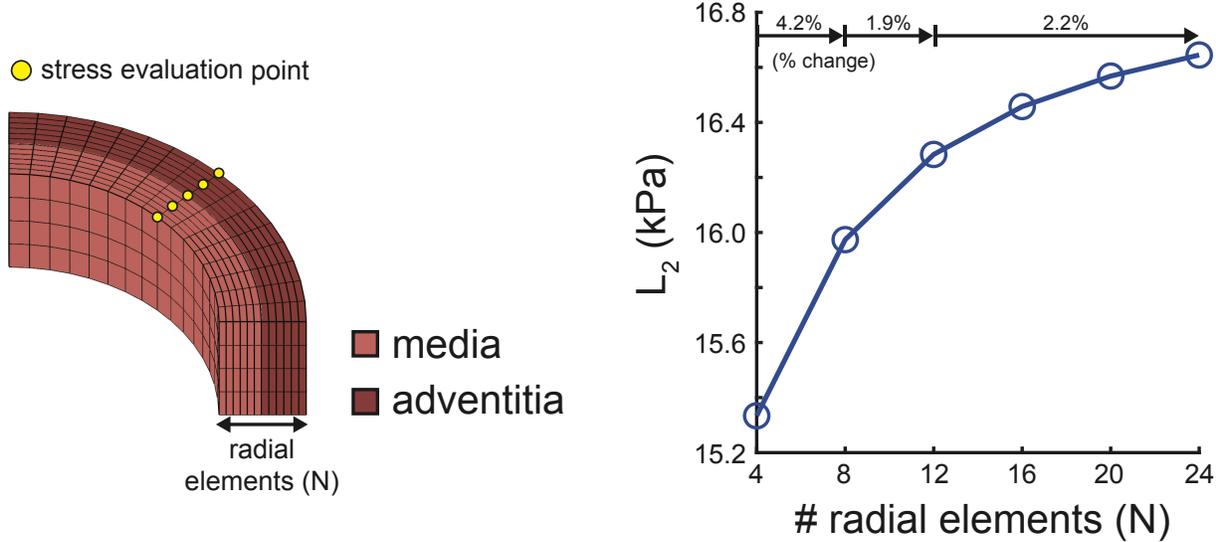

Mesh convergence study for the idealized artery model. The 2-norm criteria ($L^2$) for the 1st principal stress were evaluated at corresponding positions for radial mesh densities of 4, 8, 12, 16, 20, and 24 elements (equal number of elements in each layer). A change from 12 to 24 radial elements resulted in a change in $L_2$ for the 1st principal stress of <2.5%.

## Supplementary Figure 2

### Original Sampling

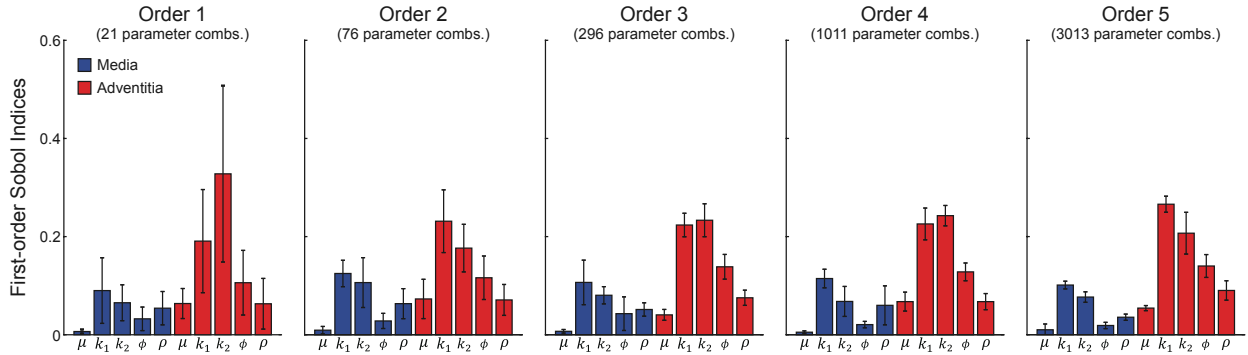

### 2x Oversampling

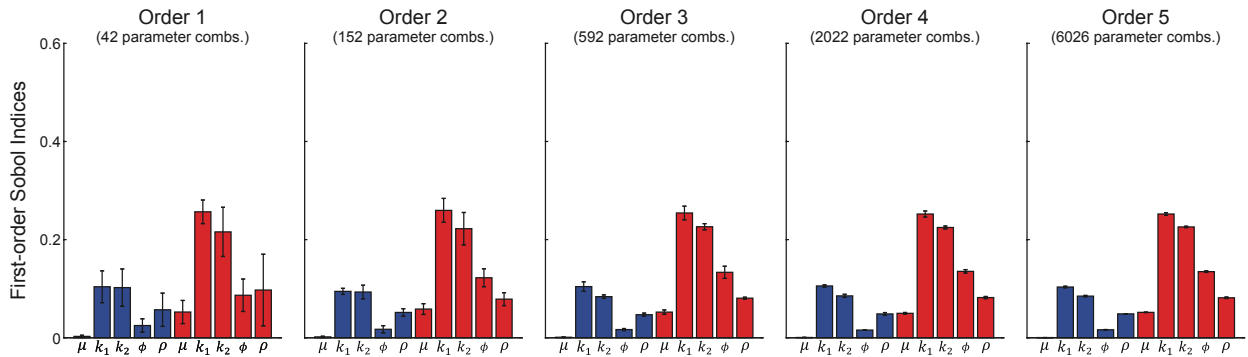

First-order Sobol indices in medial and adventitial layers at different parameter sampling rates. Original sampling (top row) and 2x oversampling (bottom row) across orders 1 through 5, with 5 runs for each order. Data are reported as mean ± standard deviation.

**Supplementary Figure 3**

**Order 2**

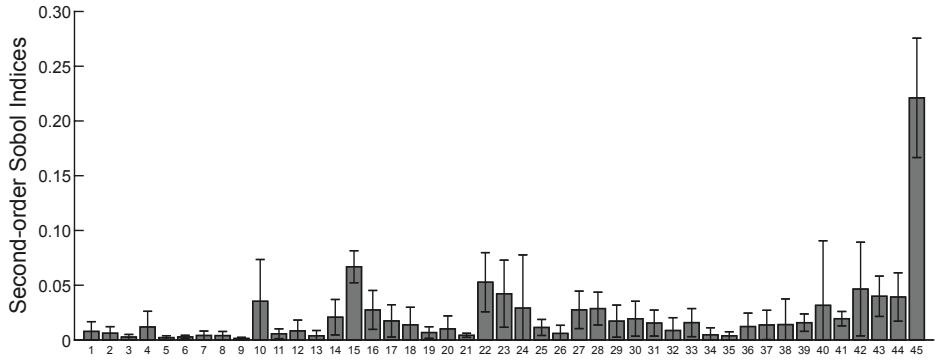

**Order 3**

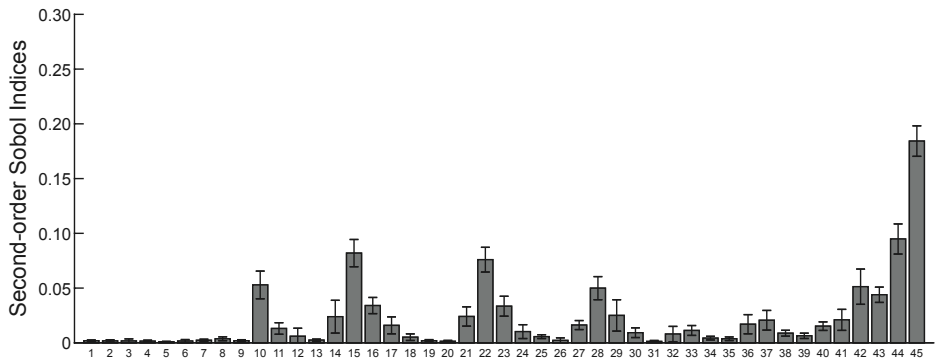

**Order 4**

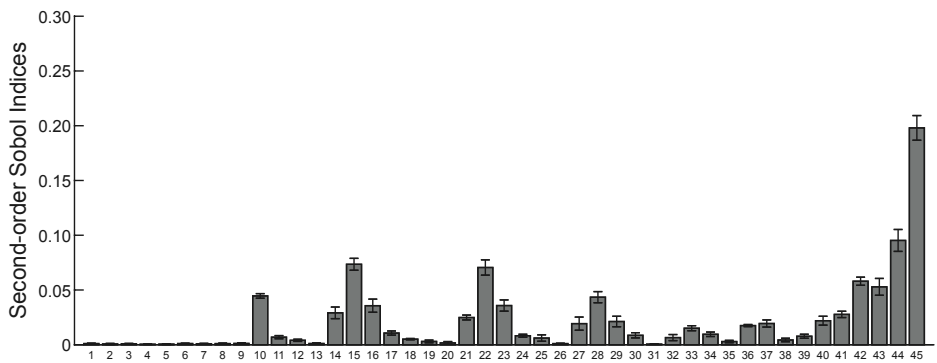

**Order 5**

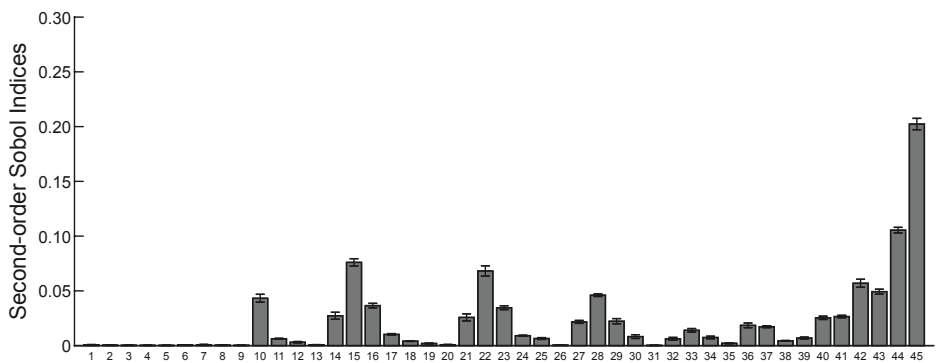

Second-order Sobol indices (i.e., binary interactions) across orders 2 through 5, with 5 runs for each order. Note that binary interactions cannot be evaluated at order 1. Data are reported as mean ± standard deviation.

X-axis legend (M: media, A: adventitia)

1: $M,\mu$ - $M,k_1$
2: $M,\mu$ - $M,k_2$
3: $M,\mu$ - $M,\rho$
4: $M,\mu$ - $M,\phi$
5: $M,\mu$ - $A,\mu$
6: $M,\mu$ - $A,k_1$
7: $M,\mu$ - $A,k_2$
8: $M,\mu$ - $M,\rho$
9: $M,\mu$ - $A,\phi$
10: $M,k_1$ - $M,k_2$
11: $M,k_1$ - $M,\rho$
12: $M,k_1$ - $M,\phi$
13: $M,k_1$ - $A,\mu$
14: $M,k_1$ - $A,k_1$
15: $M,k_1$ - $A,k_2$
16: $M,k_1$ - $A,\rho$
17: $M,k_1$ - $A,\phi$
18: $M,k_2$ - $M,\rho$
19: $M,k_2$ - $M,\phi$
20: $M,k_2$ - $A,\mu$
21: $M,k_2$ - $A,k_1$
22: $M,k_2$ - $A,k_2$
23: $M,k_2$ - $A,\rho$

24: $M,k_2$ - $A,\phi$
25: $M,\rho$ - $M,\phi$
26: $M,\rho$ - $A,\mu$
27: $M,\rho$ - $A,k_1$
28: $M,\rho$ - $A,k_2$
29: $M,\rho$ - $A,\rho$
30: $M,\rho$ - $A,\phi$
31: $M,\phi$ - $A,\mu$
32: $M,\phi$ - $A,k_1$
33: $M,\phi$ - $A,k_2$
34: $M,\phi$ - $A,\rho$
35: $M,\phi$ - $A,\phi$
36: $A,\mu$ - $A,k_1$
37: $A,\mu$ - $A,k_2$
38: $A,\mu$ - $A,\rho$
39: $A,\mu$ - $A,\phi$
40: $A,k_1$ - $A,k_2$
41: $A,k_1$ - $A,\rho$
42: $A,k_1$ - $A,\phi$
43: $A,k_2$ - $A,\rho$
44: $A,k_2$ - $A,\phi$
45: $A,\rho$ - $A,\phi$